\documentclass[review,number,sort&compress]{elsarticle}

\usepackage[utf8]{inputenc}  
\usepackage{amsmath}         
\usepackage{subfigure}       

\usepackage{pifont}    
\usepackage{natbib}    
\usepackage{geometry}  
\usepackage{fleqn}     
\usepackage{graphicx}  
\usepackage{txfonts}   
\usepackage{hyperref}  
\usepackage{dcolumn}
\newcolumntype{.}{D{.}{.}{-1}}

\begin{document}

\title{Real Time Evolvable Hardware for Optimal Reconfiguration of Cusp-Like Pulse Shapers}

\author[ucm]{Juan Lanchares}
\ead{julandan@dacya.ucm.es}
\author[ucm]{Oscar Garnica\corref{cor1}}
\ead{ogarnica@dacya.ucm.es}
\author[ucm]{José L. Risco-Martín}
\ead{jlrisco@dacya.ucm.es}
\author[ucm]{J. Ignacio Hidalgo}
\ead{hidalgo@dacya.ucm.es}
\author[ucm]{José M. Colmenar}
\ead{jmcolmenar@ajz.ucm.es}
\author[ucm]{Alfredo Cuesta}
\ead{acuestai@pdi.ucm.es}

\cortext[cor1]{Corresponding author}
\address[ucm]{Facultad de Informática, Universidad Complutense de Madrid (UCM), C/Prof. José García Santesmases s/n, 28040 Madrid, Spain
           }

\begin{abstract}
The design of a cusp-like digital pulse shaper for particle energy measurements requires the definition of four parameters whose values are defined based on the nature of the shaper input signal (timing, noise, \ldots) provided by a sensor. However, after high doses of radiation, sensors degenerate and their output signals do not meet the original characteristics, which may lead to erroneous measurements of the particle energies. We present in this paper an evolvable cusp-like digital shaper, which is able to auto-recalibrate the original hardware implementation into a new design that match the original specifications under the new sensor features.
\end{abstract}


\maketitle


\section{Introduction}\label{sec:Intro}
The detector-preamplifier configuration of a common spectroscopy system produces a pulse with an initial short rise time followed by a long exponential tail. The cusp-like pulse shaper is a well-known algorithm used in spectroscopy to analyze these exponential signals \cite{Jordanov1994}. The pulse energy is measured as the peak height of the resultant shape. In this digital shaper, the exponential signal is transformed into a symmetrical shape, with the leading edge proportional to $t^2+t$ with $t$ being the time, using series of differentiators and integrators as well as a set of parameters that are defined according to the sensor and analog electronic features. However, data signal acquisition electronics are prone to degradation, basically due to extreme environmental conditions such as high radiation levels \cite{Gersch2000}, high amplitude temperature cycles of detector materials and digital electronics, non-filtered spurious peaks of power supply, etc. Several approaches have been proposed to tackle these problems, most of them focused in annealing the sensor to restore its features after high doses of radiation \cite{Eremin2002, Li2005}. In \cite{Fernandes2010}, a real time algorithm based on a trapezoidal shaper was developed and implemented in a FPGA to determine the amplitude of an exponential pulse proving the suitability of this kind of shapers to this aim. However they do not tackle the problems related with detector aging. Several algorithms have been proposed to design adaptive shapers: Least-Mean-Square (LMS), Digital Penalized LMS (DPLMS), Wiener algorithm and Discrete Fourier Transform (DFT) \cite{Oppenheim2010, Gatti2004}. More recently,~\cite{Regadio2014} studied an adaptive FIR filter where the output pulse is converted into a desired pulse by adjusting the filter coefficients in real-time.

In this paper, we develop and test an approach than can be combined with those: an evolvable cusp-like digital pulse shaper, which adapts itself to tackle unpredictable modifications in the components of the data acquisition chain. To this aim, we apply Evolvable HardWare (EHW) concepts \cite{Greenwood2006, Sekanina2004} that have been used in the past to design resilience systems~\cite{ Walker2013, Lala2006, Greenwood2005}. We have classified our design as evolvable as it better satisfies the definition given by Greenwood in~\cite{Greenwood2006} than a reconfigurable or adaptive system.

In our work, when the shaper input signal is modified due to sensor degradation, the platform is able to evolve all the parameters of the digital pulse shaper in order to obtain a symmetrical shape close to the one obtained with the original input signal and the original components in the data acquisition chain. Results show that the EHW is able to design a target cusp-like shaper defined just by the expected output signal, although the input signal has changed due to unexpected degeneration of the radiation sensors. The technique we propose in this paper is similar to that in~\cite{Lanchares2013} where the authors designed an evolvable shaper to recover the original response after sensor degradation. In ~\cite{Lanchares2013}, the prototyping platform was only tested with trapezoidal digital shapers. In this new work we present the design of a new kind of shaper using the same prototyping platform, and thus validating both the platform architecture and the Genetic Algorithm. In addition, three different fitness functions have been also tested to validate the convergence of the searching algorithm, and the behavior of the algorithm for the new shaper is compared with previous results in literature.  As a result, this work presents a new evolvable filter, results regarding the new fitness functions, and extends the method and confirms its feasibility as a procedure to adapt the filter behavior, independently of the target digital filter.

This paper is organized as follows: Section~\ref{sec:shaper} presents the evolvable cusp-like shaper, Section~\ref{sec:Experiments} presents the experimental results, and finally Section~\ref{sec:Conclusions} covers the conclusions.


\section{Evolvable Cusp-Like Pulse Shaper}\label{sec:shaper}

Evolvable hardware is hardware capable of modifying its internal structure autonomously, without a human designer had foreseen the reasons that actually trigger the modification, and it is performed during circuit operation once the system has been deployed \cite{Greenwood2006}. The evolution is triggered either when hardware works erroneously or when the quality of the results dismiss~\cite{Higuchi2006}. 

A recursive algorithm that converts a digitized exponential pulse $v(n)$ into a symmetrical pulse $s(n)$ is given by \eqref{eq:Conformador01} to \eqref{eq:Conformador05} borrowed from \cite{Jordanov1994}:

\begin{eqnarray}
 d^{k}(n) & = & v(n) - v(n-k), \label{eq:Conformador01} \\
 d^{1}(n) & = & v(n) - v(n-1), \label{eq:Conformador02} \\
 p(n) & = & p(n-1) + d^{k}(n) - k \cdot d^{1}(n-l), n \geq 0, \label{eq:Conformador03} \\
 q(n) & = & q(n-1) + m_2 \cdot p(n), n \geq 0, \label{eq:Conformador04} \\
 s(n) & = & s(n-1) + q(n) + m_1 \cdot p(n), n \geq 0  \label{eq:Conformador05}
\end{eqnarray}

\noindent where $v(n)$, $p(n)$, $q(n)$ and $s(n)$ are equal to zero for $n<0$. 
In \eqref{eq:Conformador01}-\eqref{eq:Conformador05} appear four parameters --$k$, $l$, $m_1$, and $m_2$-- whose values are defined according to the characteristics of the shaper input signal $v(n)$. According to \cite{Jordanov1994}: $m_1$ and $m_2$ only depend on the decay time constant, $\tau$, of $v(n)$, and the sampling period, $T_{\mathrm{clk}}$, as given in \eqref{eq:M}; the delay parameter $l$ determines the duration of rising and falling edges; and the parameter $k$ depends on $l$ as $k = 2 \cdot l + 1$. At the functional level, the parameters $m_1$ and $m_2$ set the digital gain of the shaper.

\begin{equation}\label{eq:M}
\frac{m_1}{m_2}=\left( e^{\frac{T_{\mathrm{clk}}}{\tau}}-1 \right)^{-1}
\end{equation}

\begin{figure*}[ht]
\centering
\includegraphics[width=0.8\textwidth]{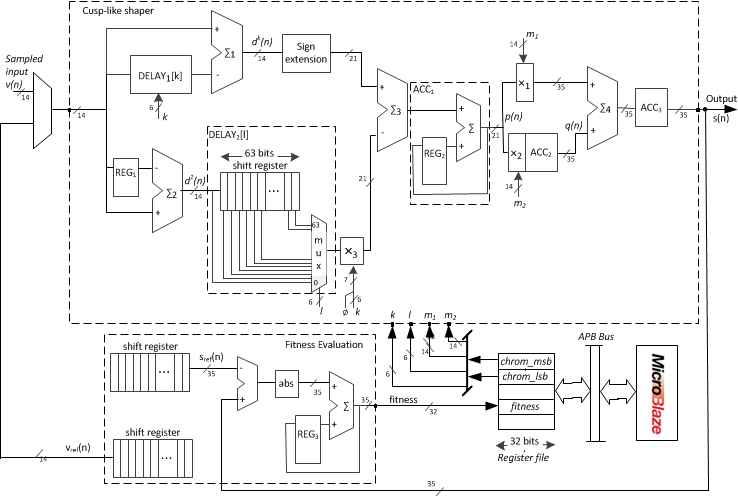}
\caption{Block diagram of system prototype. \texttt{REG}$_{\{1,2,3\}}$ are register, \texttt{DELAY}$_{\{1,2\}}$ are the delay pipelines, $\Sigma_{\{1,2,3,4\}}$ are  adders/subtractors, \texttt{ACC}$_{\{1,2,3\}}$ are accumulators and \texttt{X}$_{\{1,2,3\}}$ are multipliers. $d^k(n)$, $d^1(n)$, $p(n)$, $q(n)$, and $s(n)$ are presented in \eqref{eq:Conformador01}-\eqref{eq:Conformador05}. Signal \textit{fitness} is truncated at the output port of the Fitness Evaluation module. For the sake of clarity, control module and control signals have been obviated in this figure.}
\label{fig:BlocksDiagram}       
\end{figure*}

Fig.~\ref{fig:BlocksDiagram} illustrates the block diagram of our prototype that includes the configurable cusp-like digital shaper. It is a variant of the original one proposed in \cite{Jordanov1994}, and similarly to that it is configurable with four parameters, $(k,l,m_1,m_2)$, that in our design can be modified during circuit operation by the $\mu$P. This prototype has been implemented on a Virtex-6 ML-605 Evaluation Kit. It works at 50 MHz and occupies 1658 out of 37680 (4\%) slices in the Virtex-6 XC6VLX240T-1FFG1156 included in the evaluation kit.

The shaper has three multipliers and two delay pipelines. Both pipelines are configured with 6-bit wide parameters $k$ and $l$, respectively, that set the actual delay of these blocks to a value in the range of allowed delays $\left[ 0,63 \right]$. In addition, $k$ is converted from 6-bit binary to $7$-bit two's complement representation because multiplier $\mathrm{X}_3$ uses it as input value as well. The same principles apply to $m_1$ and $m_2$, where 14-bit two's complement multiplication is performed  because of the input data bus size. 

This cusp-like digital shaper can be customized during circuit operation to implement any of the $2^{(6+6+14+14)} \approx 10^{12}$  different shapers that can be defined using the four parameters. The reconfiguration is triggered by the Fitness Evaluation Module.

\subsection{Fitness Evaluation Module}\label{sec:fitness}

This module uses the configurable cusp-like digital shaper to evaluate an instantiation of the shaper defined by the four parameters, $(k,l,m_1,m_2)$, received from the $\mu$P. To this end, this module generates a reference input test vector, $v_{\mathrm{ref}}(n)$, that clones a typical signal generated by the sensor, and sends it to the shaper\footnote{In future versions, $v(n)$ will directly come from the sensor.}. The shaper generates the output vector, $s(n)$, that is sent back to the evaluation module that computes the fitness of the current shaper. 

The reference output signal, $s_\mathrm{ref}(n)$, has been previously calculated as the output of the reference cusp-like shaper, defined by the four parameters $(k,l,m_1,m_2)_{\mathrm{ref}}$, when receives the reference input test vector, $v_\mathrm{ref}(n)$. In this work, we have tested three fitness functions, namely $F_1$, $F_2$, and $F_3$. The first one is:

\begin{equation}\label{eq:F1}
F_1 = \lvert \max_{n=0 \ldots N-1}{s(n)} - \max_{n=0 \ldots N-1}{s_{\mathrm{ref}}(n)} \rvert
\end{equation}

where $\max_{n=0 \ldots N-1}{s(n)}$ and $\max_{n=0 \ldots N-1}{s_{\mathrm{ref}}(n)}$ are the maximum heights of the actual output, $s(n)$, provided by the configurable shaper, and the desired reference output, $s_\mathrm{ref}(n)$, respectively. The second one is:

\begin{equation}\label{eq:F2}
F_2 = \sum_{n=0}^{N-1} \lvert s(n) - s_\mathrm{ref}(n) \rvert
\end{equation}

that is the cumulative error between the two signals. Finally, the third one is:

\begin{eqnarray}
 F_3 & = & F_1 + F_2 \nonumber \\
     & = & \lvert \max_{n=0 \ldots N-1}{s(n)} - \max_{n=0 \ldots N-1}{s_{\mathrm{ref}}(n)} \rvert + \sum_{n=0}^{N-1} \lvert s(n) - s_\mathrm{ref}(n) \rvert
\end{eqnarray}

that simultaneously takes into account the peak height of the signal and the cumulative error.

\begin{figure}[ht]
\centering
\subfigure[]{
    \includegraphics[width=0.75\columnwidth]{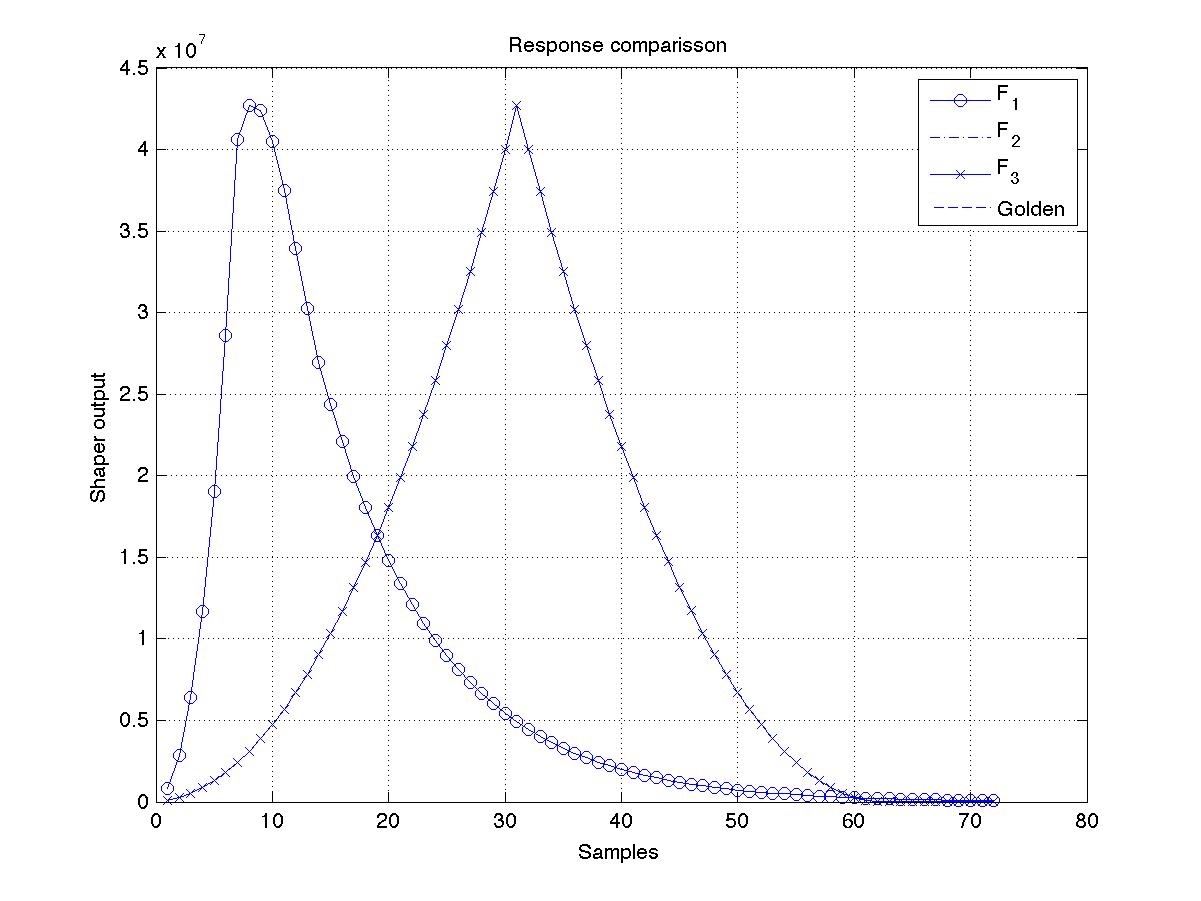}
    \label{fig:fitness_comparison_scratch}
}
\subfigure[]{
    \includegraphics[width=0.75\columnwidth]{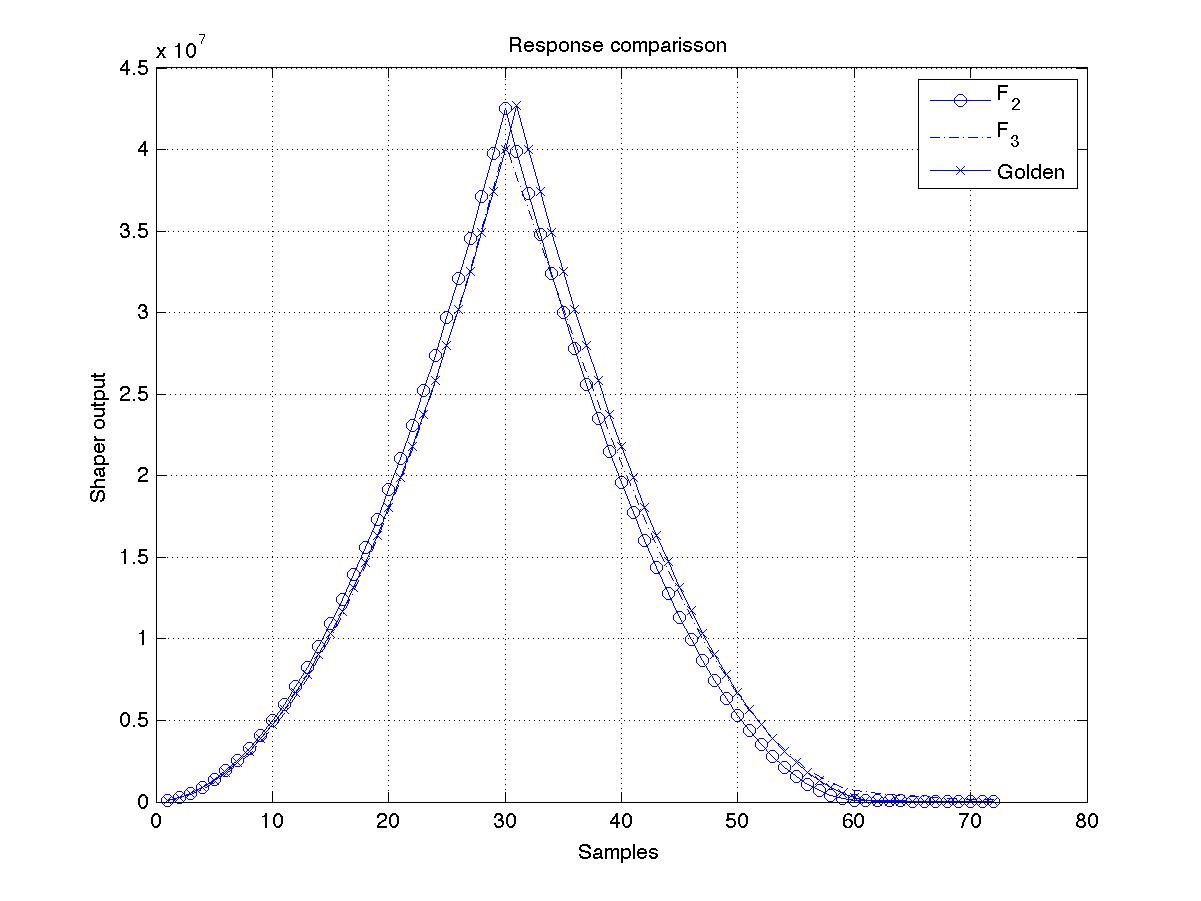}
    \label{fig:fitness_comparison_degenerated}
}
\caption[]{Regenerated response for the three fitness functions, $F_1$, $F_2$, $F_3$. For comparison purposes we also present the golden response. \subref{fig:fitness_comparison_scratch} Regenerated response starting from scratch. \subref{fig:fitness_comparison_degenerated} Regenerated response after signal degeneration.}
\label{fig:fitness_comparison}
\end{figure}

In  Figure~\ref{fig:fitness_comparison} we present some of the results obtained with these fitness functions. In particular, in Figure~\ref{fig:fitness_comparison_scratch} we present the results obtained by the three fitness functions for the worst-case scenario, that is, when the evolution aim is to recover reference signal starting from scratch (i.e. from a random chromosome).  The results with $F_2$ are identical to $F_3$, and both coincide with $s_\mathrm{ref}(n)$. Therefore, both of them reach the best chromosome. On the other hand, the output signal with $F_1$ is shifted to the left and is asymmetrical, when one of the cusp-like features for better noise filtering is that the rising and falling edges should be symmetrical. This a common behavior that we consistently observe in all of our experiments. Hence, we focused the studies on the fitness functions $F_2$ and $F_3$.

Figure~\ref{fig:fitness_comparison_degenerated} presents the results for $F_2$ and $F_3$ when time constant of $v(n)$ has been degenerated from $200~\mu\mathrm{s}$ to $140~\mu\mathrm{s}$. Again, $F_2$ results are very similar to $F_3$ ones.  So, other factors have to be studied in order to make a decision about which fitness function has to be used.  In the case of $F_2$, the average number of generations is 114, being the minimum number of generations 7 and the maximum 1156. The average time is 0.063 minutes, being the minimum 0.004 minutes and the maximum 0.63 minutes.
For $F_3$, the average number of generations is 2474, being the minimum number of generations 19 and the maximum 16165. The average time is 1.40 minutes, being the minimum 0.01 minutes and the maximum 9.16 minutes. These data have been obtained after 150 runs for each fitness function.
Regarding hardware resources, in the design with $F_2$ the number of flips flops and LUTs are 7471 and 8327, respectively. These figures are slightly higher for $F_3$, 7573 flips flops and 8676 LUT.

According to these results, we are going to use $F_2$ as the fitness function in our experiments.

\subsection{MicroBlaze Processor}\label{sec:microblaze}

MicroBlaze is one of the $\mu$P that Xilinx provides in its IP libraries. It will store the population of individuals in the Genetic Algorithm (GA) and will execute the following tasks of the GA: population initialization, selection of the individuals, crossover and mutation operators, and checks the finish criteria. Although performing all these tasks, this module acts as a slave in the communication with the evolvable shaper. The evolvable shaper will conduct the optimization process, indicating when the GA should be run and requesting the individual to be evaluated in order to find their fitness value.

\subsection{Evolutionary Algorithm}

We have implemented a standard GA~\cite{Goldberg1989} which evolves a population of 125 individuals where each individual encodes the values of the four parameters of the shaper, $(k,l,m_1,m_2)$, using a binary representation. Therefore the GA evolves a population of 125 shaper design representations trying to find the best one.

\begin{figure}
\centering
\includegraphics[width=1.0\columnwidth]{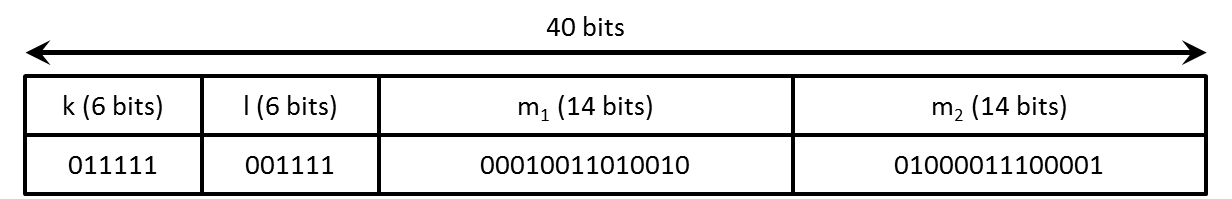}    
\caption{Individual encoding.}
\label{fig:IndividualEncoding}
\end{figure}

As exposed above, 40 bits, $6+6+14+14$, are required to codify the four parameters. Therefore, an individual uses 40 bits to encode a solution. Fig.~\ref{fig:IndividualEncoding} depicts an example of an individual which encodes the shaper with parameters $(k,l,m_1,m_2)=(31,15,1234,4321)$.

Our implementation of the GA has the particularity that the different operators are executed in different modules of the system. Thus, first, an initial population is randomly generated in the $\mu$P. Next, individuals are sequentially requested by the shaper module, where each one is evaluated, and sent back to the $\mu$P. Once the population is fully evaluated, $\mu$P applies elitism and the four best individuals are copied into the offspring population. The offspring is completed with new individuals generated using binary tournament selection and a 1-point crossover operator. Finally, the mutation operator randomly selects, according to mutation probability, one bit in the chromosome and inverts its state. We keep the size of the elite small because if not, the algorithm is usually trapped in local optima.


\section{Experiments}\label{sec:Experiments}

Sensor degradation has two consequences. On the one hand the peak voltage of the events at the sensor output is reduced \cite{Gersch2000,Singh1968}, and in the other, the leakage current is increased and, consequently, increases the noises (both serial and parallel) at the output of the preamplifier behind the sensor \cite{Medina2012}. 

In  a first set of experiments we check the reliability of the method by studying the number of successful evolutions and their running times for the worst-case scenario when the shaper parameters has to be set starting from scratch --a randomly generated chromosome--.  For this scenario, we have run 150 experiments starting with different initial chromosomes. In all the cases, the GA found the optimal solution. As stated in Section~\ref{sec:fitness}, the average number of generations was 114, being the minimum number of generations 7 and the maximum 1156. The average time was 0.063 minutes, being the minimum 0.004 minutes and the maximum 0.638 minutes. The average time  in~\cite{Lanchares2013} for the trapezoidal shaper was 22.32 minutes, being the minimum 0.119 minutes and the maximum exceeds 3 hours. Table~\ref{tab:comparison_convergence_times} presents the detailed comparison of the convergence times, that are between one and three orders of magnitude lower than for the trapezoidal shaper. This indicates that the search space of cus-like shapers is likely to have a less number of local minimums than the search space of the trapezoidal one, given that the number of  parameters to evolve is the same in both cases. According to this comparison, the convergence times are strongly-dependent on the shaper kind, and this is a new constraint to be taken into account in order to select the most suitable evolvable shaper for a given application.

\begin{table}
\centering
\begin{tabular}{|c|. .|}
\hline
\textbf{} & \multicolumn{1} {c}{Cusp} &  \multicolumn{1}{c|} {Trapezoidal~\cite{Lanchares2013}} \\
\hline
Best exec.                 &  0.24   & 7.14 \\
\hline
Worst exec.                &  38.28  & 11428 \\
\hline
Average                    &  3.78   & 1339  \\
\hline
$1^{\textrm{st}}$ quartil  &  1.69   & 90    \\
\hline
$2^{\textrm{nd}}$ quartil  &  2.48   & 1064  \\
\hline
$3^{\textrm{rd}}$ quartil  &  4.13   & 2320  \\
\hline
\end{tabular}
\caption{Comparison of convergence times (in seconds) of the GA for two different shaper types.}\label{tab:comparison_convergence_times}
\end{table}

\begin{figure}[ht]
\centering
\subfigure[]{
    \includegraphics[width=0.75\columnwidth]{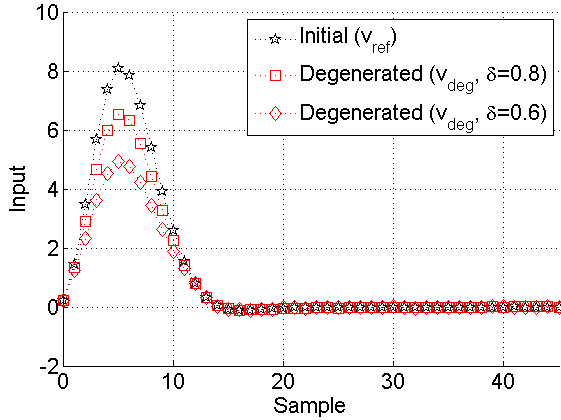}
    \label{fig:ThreeInputs}
}
\subfigure[]{
    \includegraphics[width=0.46\columnwidth]{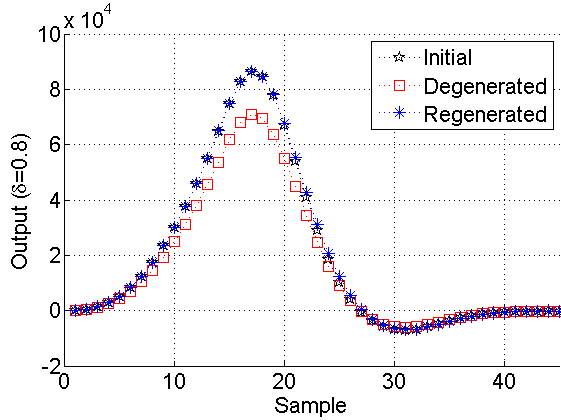}
    \label{fig:ThreeOutputs80}
}
\subfigure[]{
    \includegraphics[width=0.46\columnwidth]{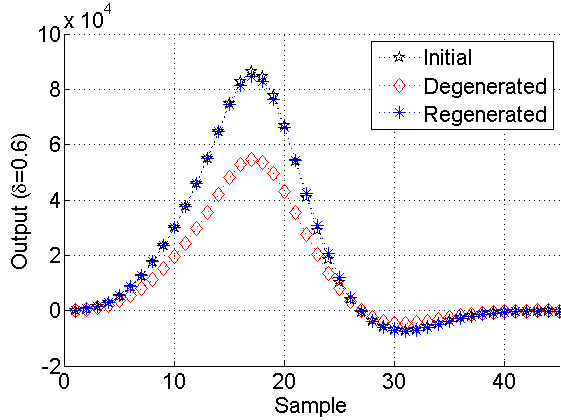}
    \label{fig:ThreeOutputs60}
}
\caption[]{\subref{fig:ThreeInputs} Initial and two degenerated events with $\delta = 0.8$ and $\delta = 0.6$, respectively. Two EHW recalibrations of the original shaper: \subref{fig:ThreeOutputs80} For the first degenerated event $(k,l,m_1,m_2)_{\mathrm{reg}}=(31,15,68,16)$, and \subref{fig:ThreeOutputs60} for the second degenerated event $(k,l,m_1,m_2)_{\mathrm{reg}}=(31,15,89,20)$.}
\label{fig:FinalPlots}
\end{figure}

In a second set of experiments we have tested our evolvable shaper using real data from the Castilla-La Mancha Neutron Monitor (CaLMa) located in the Science and Technology Park of Guadalajara (GuadaLab). We have modeled sensor degeneration as follows: (i) we have attenuated the initial signal of Fig.~\ref{fig:ThreeInputs} $(v_{\mathrm{ref}})$, and (ii) we have added serial $\kappa^s(n)$ and parallel $\kappa^p(n)$ noises to the attenuated signal as follows:

\begin{equation}\label{eq:degeneration}
v_{\delta}(n) = \left\{ \begin{array}{rl}
\delta \cdot v(n) + (1-\delta) + \kappa^s(n) + \kappa^p(n)  & \mbox{if } v(n) > 1 \\
v(n)   & \mbox{otherwise}\\
       \end{array} \right.
\end{equation}


\noindent where $v(n)$ is the value of the original signal, $\delta$ is the scaling parameter, and $v_{\delta}(n)$ is the value of the degenerated signal, $\kappa^s(n)$ is white Gaussian noise and  $\kappa^p(n)$ is $\frac{1}{f^2}$ noise.

Fig.~\ref{fig:ThreeInputs} shows the event, the one out of the 10672 events in CaLMA data, that will be used as reference signal, as well as the two degenerated signals after applying \eqref{eq:degeneration} for $\delta=0.8$ and $\delta=0.6$.

 
The original cusp-like shaper has $(k,l,m_1,m_2)_\mathrm{ref}=(31,15,57,13)$. The response of this shaper for the original and degenerated signals is illustrated in Fig.~\ref{fig:ThreeOutputs80} and Fig.~\ref{fig:ThreeOutputs60} for $\delta = 0.8$ and $\delta = 0.6$, respectively. After $150$ tests for each experiment, and $4.16 \pm 2.54$ seconds on average, the new parameters were obtained, $(k,l,m_1,m_2)_{\mathrm{reg}}=(31,15,68,16)$ for $\delta=0.8$ and $(k,l,m_1,m_2)_{\mathrm{reg}}=(31,15,89,20)$ for $\delta=0.6$. The responses for the new shapers are illustrated in Fig.~\ref{fig:ThreeOutputs80} and \ref{fig:ThreeOutputs60} with label \textit{Regenerated}. Both of them depict the level of optimization that our EHW prototype performs with cumulative errors $F_2=18341.5$ and $F_2=22341.1$, respectively. As both figures show, the new shape is quite similar to the original one. Thus, after sensors degradation, degenerated events can be clearly detected. In addition, all the input data are scaled to the bus width, to gain in precision. However, for better clarity we have re-scaled all the figures and tables to the original input interval.

\begin{figure}[ht]
\centering
\subfigure[]{
    \includegraphics[width=0.85\columnwidth]{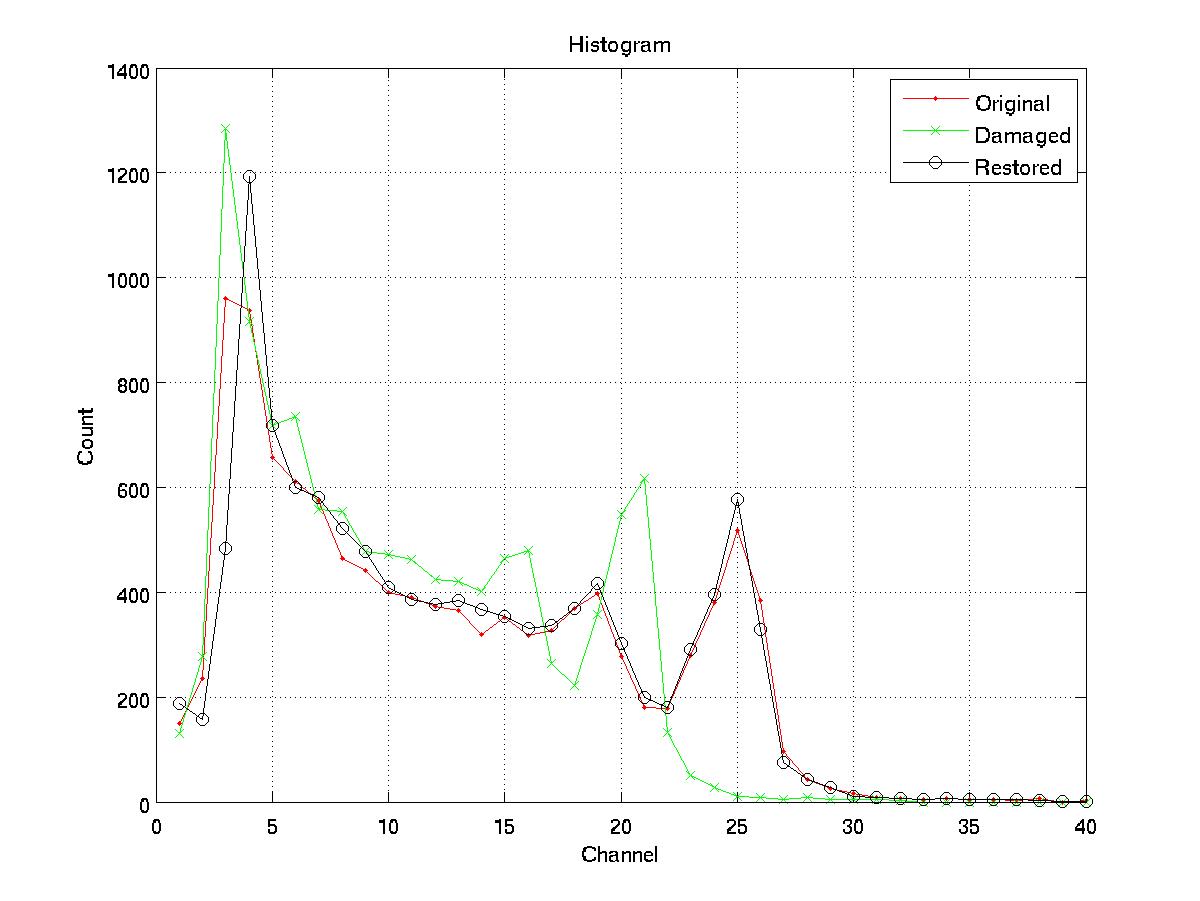}
    \label{fig:Histogram80}
}
\subfigure[]{
    \includegraphics[width=0.85\columnwidth]{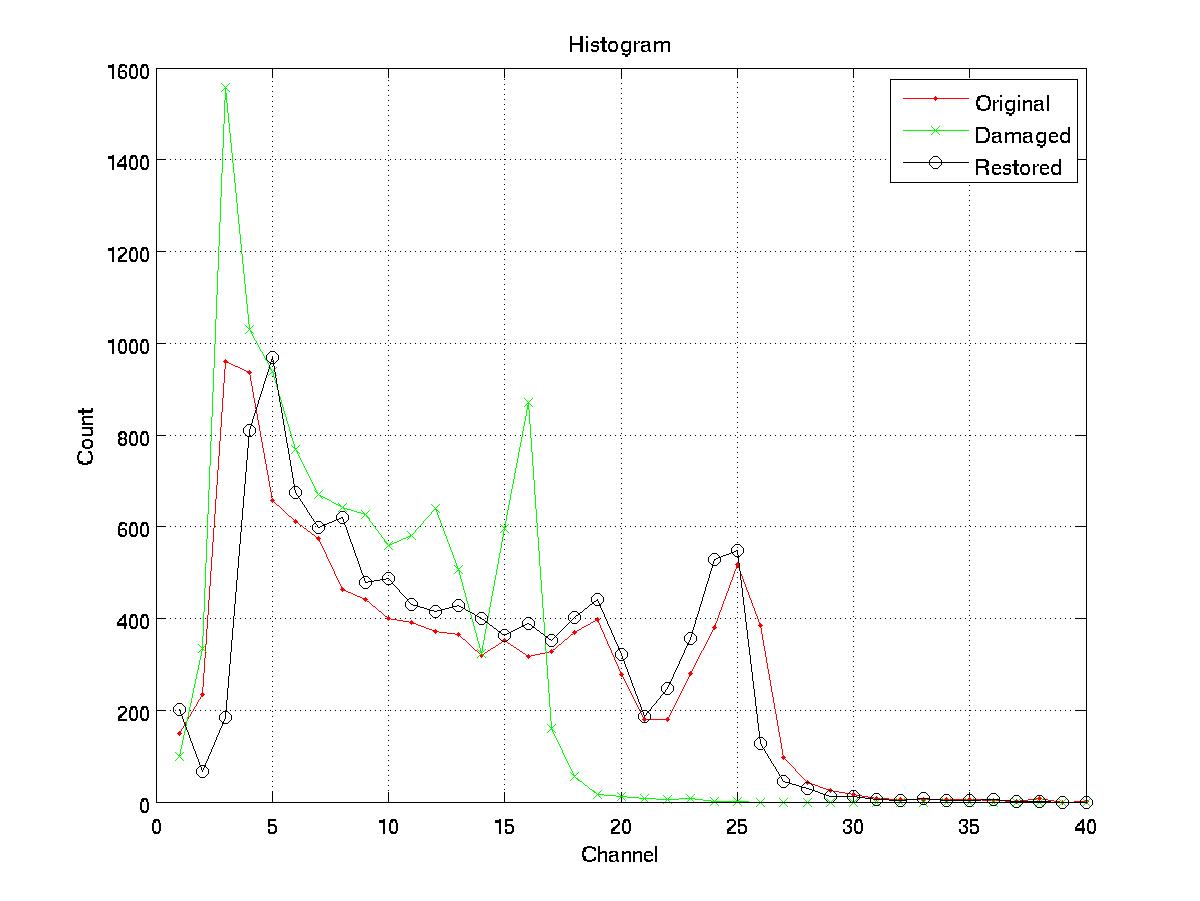}
    \label{fig:Histogram60}
}
\caption[]{Histograms of the original and degenerated sensor signals with \subref{fig:Histogram80} $\delta=0.8$ and \subref{fig:Histogram60} $\delta=0.6$, respectively. Original: is the histogram of the original signal for shaper's parameters $(k,l,m_1,m_2)_{\mathrm{ref}}=(31,15,57,13)$. Damaged is the histogram of the degraded signal for the same parameters. Restored is the histogram of the degraded signal for optimized shapers' parameters $(k,l,m_1,m_2)_{\mathrm{reg}}=(31,15,68,16)$ and $(k,l,m_1,m_2)_{\mathrm{reg}}=(31,15,89,20)$, respectively.}
\label{fig:FinalHistograms}
\end{figure}

Fig.~\ref{fig:FinalHistograms} shows the overall impact of the optimized shapers in the whole real data set, measured during 40 minutes to reach up to 592.2M samples and 10672 events. The number of samples and events is limited, just the required to obtain a peak in the histograms to validate the result of our optimization. As Fig.~\ref{fig:FinalHistograms} shows, the region of interest is located around triangle height $=3.5 \times 10^7$. All the events were degenerated with attenuation factors of $\delta=0.8$ and $\delta=0.6$, respectively. Thus, both histograms were shifted to the left. In Fig.~\ref{fig:FinalHistograms}, data coming from the non-degenerated sensor are labeled as \textit{Original} and the degenerated signal is labeled as \textit{Damaged}. Both outputs of the optimized shapers in the histogram are labeled as \textit{Restored}. As can be seen, the region of interest is restored to the original position.

\begin{figure}
\centering
\includegraphics[width=0.85\columnwidth]{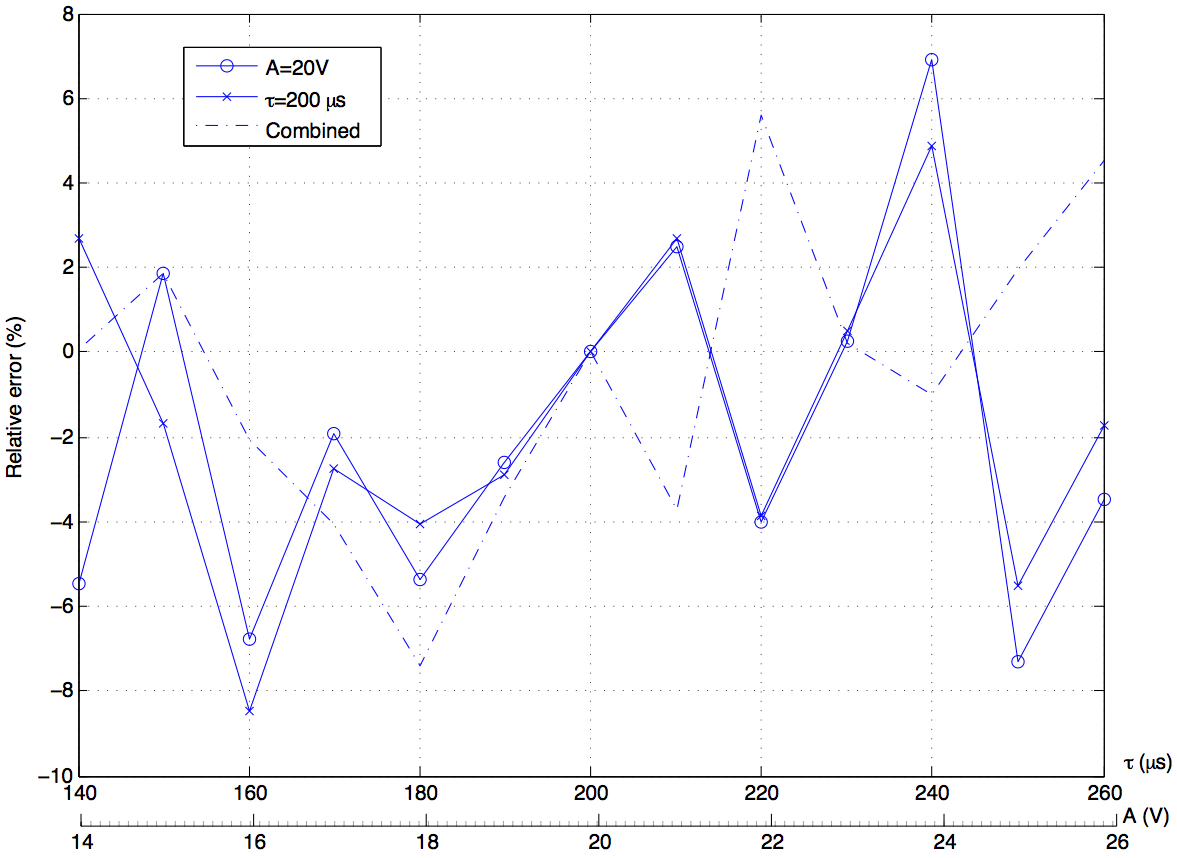}
\caption{Relative error of the peak amplitude of the CUSP-like filter output. $A=20$ represents the error for variations in $\tau$ with amplitude taking the value 20 V. $\tau=250 \mu s$ represents the error for variations in $A$ with constant time taking the value $200 \mu s$. At last, Combined represents the error for entangled variations in A and $\tau$ using the values in their corresponding axes.}
\label{fig:relative_error}
\end{figure}

In a third set of experiments, the effectiveness of our evolvable cusp-like digital shaper is tested using a suite of synthetic degenerated inputs where the level of degeneration is much higher than in the previous experiment. To this end, a reference input $v_{\mathrm{ref}}(n)=A_{\mathrm{ref}}\cdot e ^{-\frac{t}{\tau_{\mathrm{ref}}}}$ and cusp-like shaper $(k,l,m_1,m_2)_{\mathrm{ref}} = (63,31,19,2)$ have been taken, being $A_{\mathrm{ref}}=20~\mathrm{V}$, $\tau_{\mathrm{ref}}=200~\mu\mathrm{s}$, $T_{\mathrm{clk}} = 20~\mu \mathrm{s}$, and $N=72$. Next, the reference input is degraded introducing variations in amplitude $(A)$, the time constant $(\tau)$, and in both parameters simultaneously. Fig.~\ref{fig:relative_error} illustrates the results. We have measured the relative error of the peak value of the restored filter when compared to the reference output for the three experiments. X-axes represent the variations in $A$ and $\tau$. In all the experiments the GA found the optimal solution, converging in a reliable way.

The relative error never exceeds the 8\% in absolute value, although the value of  A and $\tau$ have been modified up to 30\%. As expected, error relative has a Gaussian behavior --tested using the Lilliefors test-- with value $e=-1.36 \pm 3.76\%$. In addition, the magnitude of the relative error seems to be uncorrelated with the magnitude of the signal degradation in amplitude: linear fitting slope with 95\% confidence bounds is $m = 0.1295 \pm 0.6215$ that is statistically indistinguishable of zero, the correlation coefficient $r= 0.137$, and the hypothesis testing of no correlation between these variables returns that the correlation is not significant. 

The same conclusions apply for the other two relations under study: uncorrelated with the the constant time ( $m = 0.0194 \pm 0.0692$, $r= 0.1826$, and correlation is not significant), or with both ( $m = 0.03855 \pm 0.0561$, $r = 0.415$, and correlation is not significant).



\section{Conclusions}\label{sec:Conclusions}

We have presented a prototype of an evolvable cusp-like digital shaper using concepts and ideas from evolvable hardware and evolutionary computation. Our evolvable digital shaper has been implemented in a FPGA. We have proved that this design is able to reach optimal designs under fluctuations in the input signal due to aging effect in sensor electronic. Hence, under input signal degradation, our digital shaper may automatically optimize its parameters to reach the same output signal as the one obtained under the reference input signal.

Our cusp-like evolvable shaper has been validated using a singular degeneration in real data. We have also conducted experiments
over synthetic data to validate the scalability and resilience of the digital shaper. Results show that the shaper is reconfigured in less than 1 minute, and that the level of improvement is higher as the degeneration is increased. The evolution time for the cusp-like shaper is 10 times lower that for the trapezoidal shaper and, consequently, we have proved that the evolution times are strongly-dependant of the type of filter, although the optimal configuration is attained always.

Finally, we have studied three fitness functions for the GA, and we have concluded which is the best one in terms of convergence times and hardware resources.

In summary, this work presents: a new evolvable filter; results regarding new  fitness functions; expands the validity of the method for new filters, and confirms its feasibility as a procedure to adapt the filter behavior, independently of the target digital filter.


\section*{Acknowledgements}
This work has been supported by the Spanish Government Research grants TIN 2008/00508 and 
MEC Consolider Ingenio CSD00C-07-20811. 
We would like to thank the reviewers for providing encouraging feedback and 
insightful comments that improved the content and the quality of the paper, as well as the
Castilla-La Mancha Neutron Monitor (CaLMa) experiment which provided the data used in
this paper.

\bibliographystyle{elsarticle-num}
\bibliography{biblio}

\end{document}